\documentclass[11pt]{article}
\usepackage{geometry}
\usepackage{graphics}
\newcommand{\diff}[2]{\frac{\partial #1}{\partial #2}}
\begin{document}
\title{A potential including Heaviside function\\ in 1+1 dimensional hydrodynamics by Landau\\
 {\large Its basic properties and application to data at RHIC energies}}
\author{$^{1,}$\thanks{mizoguti@toba-cmt.ac.jp} T.~Mizoguchi, $^2$H.~Miyazawa and $^3$M.~Biyajima\\
{\small $^1$Toba National College of Maritime Technology, Toba 517-8501, Japan}\\
{\small $^2$Graduate School of Science and Technology, Shinshu University, Matsumoto 390-8621, Japan}\\
{\small $^3$School of General Education, Shinshu University Matsumoto 390-8621, Japan}
}
\maketitle

\begin{abstract}
In 1+1 dimensional hydrodynamics originally proposed by Landau, we derive a new potential and distribution function including Heaviside function and investigate its mathematical and physical properties. Using the original distribution derived by Landau, a distribution function found by Srivastava {\it et al.}, our distribution function, and the Gaussian distribution proposed by Carruthers {\it et al.}, we analyze the data of the rapidity distribution on charged pions and K mesons at RHIC energies ( $\sqrt{s_{\tiny NN}} =$ 62.4 GeV and 200 GeV ). Three distributions derived from the hydrodynamics show almost the same chi-squared values provided the CERN MINUIT is used. We know that our calculations of hadron's distribution do not strongly depend on the range of integration of fluid rapidity, contrary to that of Srivastava {\it et al.} Finally the roles of the Heaviside function in concrete analyses of data are investigated.
\end{abstract}


\section{Introduction}
\label{intro}
Recently authors of ref.~\cite{Nagy:2007xn,Wong:2008ta} have mentioned a paper by Srivastava {\it et al.}~\cite{Srivastava:1992cg} as one of boost non-invariant solutions in $1+1$ dimensional hydrodynamics proposed by Landau~\cite{Landau:1953gs,Khalatnikov:1954,Belenkij:1956cd}. It is well know that the boost invariant distribution function have been proposed by Hwa~\cite{Hwa:1974gn} and independently Bjorken~\cite{Bjorken:1982qr}. It is also worthwhile to notice recently interesting papers on $1+1$ dimensional hydrodynamics published in refs.~\cite{Bialas:2007iu,Pratt:2008jj}. See also ref.~\cite{Sarkisyan:2005rt} in which the usefulness of the hydrodynamics for the collisions (proton-proton and nuclei-nuclei) is stressed.

In the present study, in particular we are interested in the paper by Srivastava {\it et al.}, because their potential and the distribution function are similar to that calculated by Landau and coauthors. Our purpose is to obtain a new solution by use of different boundary conditions in the same partially differential equation obtained by Landau. (See eq.~(\ref{eq_08}) below)

The hydrodynamics is briefly described in the following: The energy density $e$, the pressure $p$, the four velocity $u^{\mu}$, and the entropy density $s$ are necessary physical quantities. Then the conservation law of the energy-momentum and that of the entropy are given as 
\begin{eqnarray}
\diff{T^{\mu \nu}}{x^{\mu}} = 0,\quad \diff{(su^{\mu})}{x^{\mu}} = 0, 
\label{eq_01}
\end{eqnarray}
where $T^{\mu \nu} = (e+p)u^{\mu}u^{\nu} - pg^{\mu \nu}$. Introducing the temperature of the fluid $T$, the equation of state $p = c_s^2e$ with the velocity of sound $c_s$, the thermodynamical relations $de = Tds$, and $e+p=Ts$, we have the relation $dp = sdT$. Moreover, introducing the rapidity of fluid $y$ and $x^{\mu} = (t,\ x) = (\tau \cosh \eta,\ \tau \sinh \eta)$ with $\eta = \tanh^{-1}(x/t)$, $1+1$ dimensional hydrodynamics in the perfect fluid is described as follows
\begin{eqnarray}
&& \diff{}{t}(T\sinh y) + \diff{}{x}(T\cosh y) = 0,
\label{eq_02}\\
&& \left(\diff{T}{t} + c_s^2\diff{y}{x}\right)\cosh y
+ \left(\diff{T}{x} + c_s^2\diff{y}{t}\right)\sinh y = 0.
\label{eq_03}
\end{eqnarray}
Utilizing the Legendre transformation $d\chi = d(\phi +tT\cosh y - xT\sinh y)$ from the variables $(t,\ x)$ to $(T,\ y)$ with potentials $\phi$ and $\chi$, we obtain the following expressions:
\begin{eqnarray}
&& t = \diff{\chi}{T}\cosh y - \frac 1T\diff{\chi}{y}\sinh y,
\label{eq_04}\\
&& x = \diff{\chi}{T}\sinh y - \frac 1T\diff{\chi}{y}\cosh y.
\label{eq_05}
\end{eqnarray}
Moreover, we have the following relation between $y$ and $\eta$:
\begin{eqnarray}
\tau^2 &=& t^2 - x^2 = \left(\diff{\chi}{T}\right)^2 
- \left(\frac 1T\diff{\chi}{y}\right)^2, 
\label{eq_06}\\ 
y &=& \tanh^{-1}\left(\frac xt\right) 
+ \tanh^{-1}\left(\frac{\frac 1T\diff{\chi}{y}}{\diff{\chi}{T}}\right) 
= \eta + \Delta.
\label{eq_07}
\end{eqnarray}
From the above equations and relations, we obtain the following partial differential equation of the potential $\chi$ in $1+1$ dimensional hydrodynamics,
\begin{eqnarray}
\diff{^2\chi}{y^2} = c_s^2\diff{^2\chi}{\omega^2} + (1-c_s^2)\diff{\chi}{\omega}
\label{eq_08}
\end{eqnarray}
where $\omega = -\ln (T/T_0)$ (the logarithmic temperature). To look for a new solution in eq.~(\ref{eq_08}) is our aim, as mentioned before.

As the solution of eq.~(\ref{eq_08}), we have known the following two solutions~\cite{Landau:1953gs,Srivastava:1992cg}. (Notice that two different initial conditions are used.) Utilizing the method of Green function~\cite{Duffy,Namiki}, we obtain a new potential including the Heaviside function for the differential equation in 1+1 dimensional hydrodynamics. 

On the other hand, the rapidity distributions of the charged mesons at RHIC energies ($\sqrt{s_{\tiny NN}} =$ 62.4 GeV~\cite{Arsene:2008bh} and 200 GeV~\cite{Bearden:2004yx}) have been reported by BRAHMS Collaboration. We are going to analyze them by several formulae by Landau, Srivastava {\it et al.}, and ours.

In the second sec. we briefly explain the formula by Landau and the boost non-invariant formula calculated by Srivastava {\it et al.} In the third sec. we present the new formula including the Heaviside function. In the 4th one we present analyses of the data at RHIC energies by three formulae and a formula proposed by Carruthers {\it et al.}~\cite{Carruthers:1973rw} in addition to them. In the final one the concluding remarks are presented.


\section{Solutions of Potential (eq.~(\ref{eq_08})) by Landau and Srivastava et al.}
\label{solution1}
We mention the solution by Landau and that by Srivastava {\it et al.} An original formula by Landau is obtained by the following initial conditions,
\begin{eqnarray}
\left. \diff{\chi}{y}\right|_{y=0} 
= T_0le^{\omega},\quad \chi (y=\omega/c_s) = 0,
\label{eq_09}
\end{eqnarray}
where $l$ denotes the length of colliding matter.

The potential is expressed by the integral formula
\begin{eqnarray}
\chi \sim e^{-\omega} \int_{yc_s}^{\omega} e^{(1+\beta)\omega'} I_0
\left(\beta\sqrt{\omega'^2-c_s^2y^2}\right)\, d\omega' ,
\label{eq_10}
\end{eqnarray}
where $\beta = (1-c_s^2)/2c_s^2$. The entropy distribution is calculated as follows,
\begin{eqnarray}
\frac{dS}{dy} \sim \frac{dN}{dy} \sim {\rm const}\, \diff{}{\omega}
\left. \left(\chi + \diff{\chi}{\omega}\right)\right|_{\omega = \omega_f},
\label{eq_11}
\end{eqnarray}
where $\omega_f = - \ln(T_f/T_0)$. Using (\ref{eq_11}), we have the rapidity distribution
\begin{eqnarray}
\frac{dN}{dy} \sim e^{\beta \omega_f} \left[I_0(p) + \frac{\beta \omega_f}p I_1(p)\right] ,
\label{eq_12}
\end{eqnarray}
where $p= \beta\sqrt{\omega^2-c_s^2y^2}$. $I_0$ and $I_1$ are the modified Bessel functions of the 1st order and 2nd one, respectively.

\begin{figure}[htbp]
  \begin{center}
     (a)\resizebox{0.45\textwidth}{!}{\includegraphics{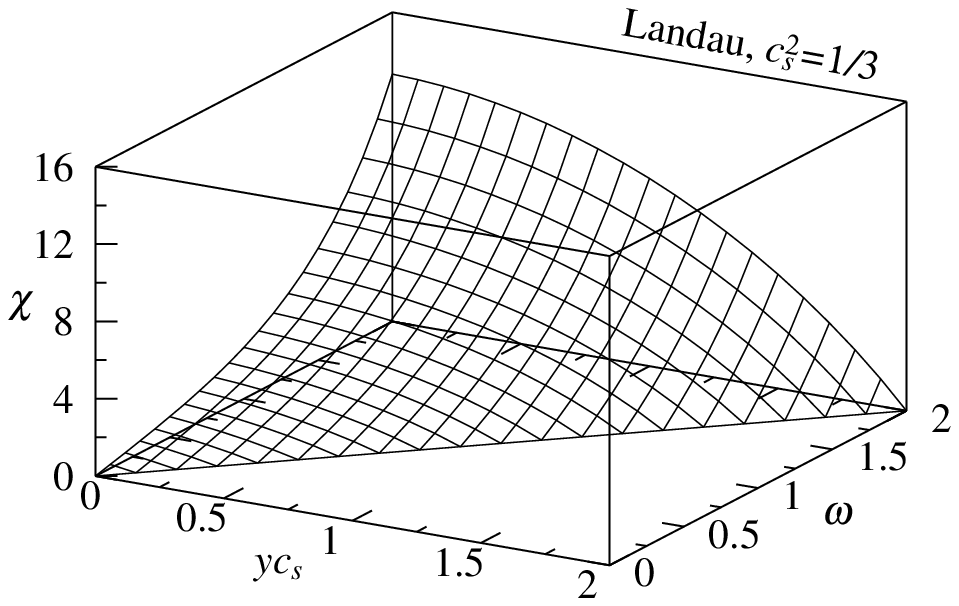}}
     (b)\resizebox{0.35\textwidth}{!}{\includegraphics{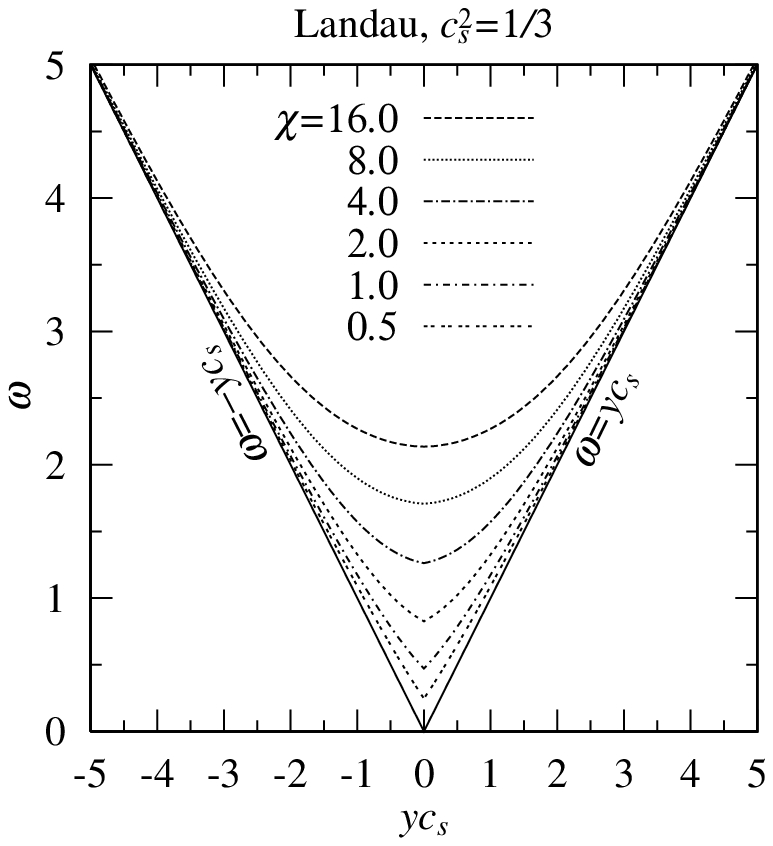}}\\
     (c)\resizebox{0.45\textwidth}{!}{\includegraphics{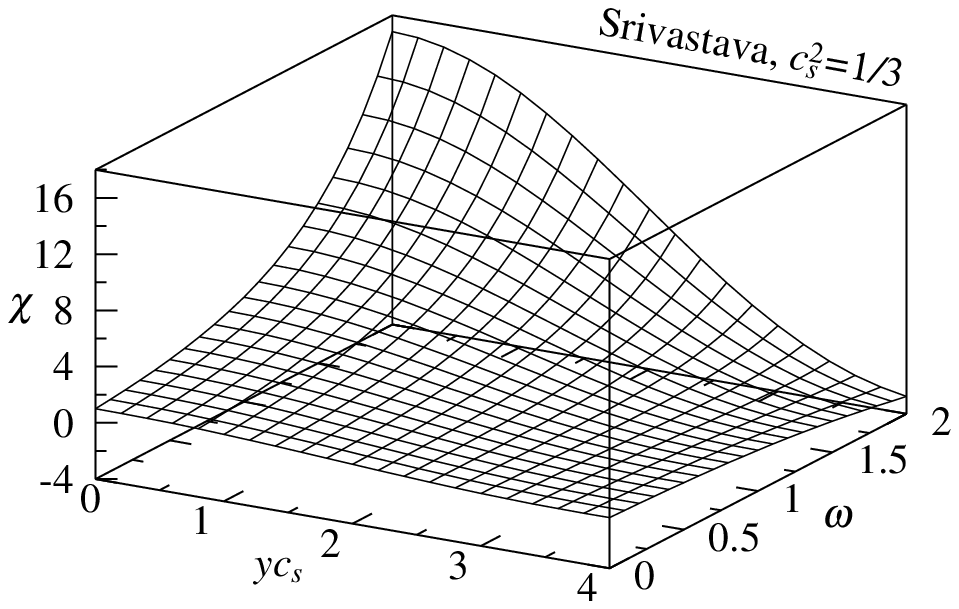}}
     (d)\resizebox{0.35\textwidth}{!}{\includegraphics{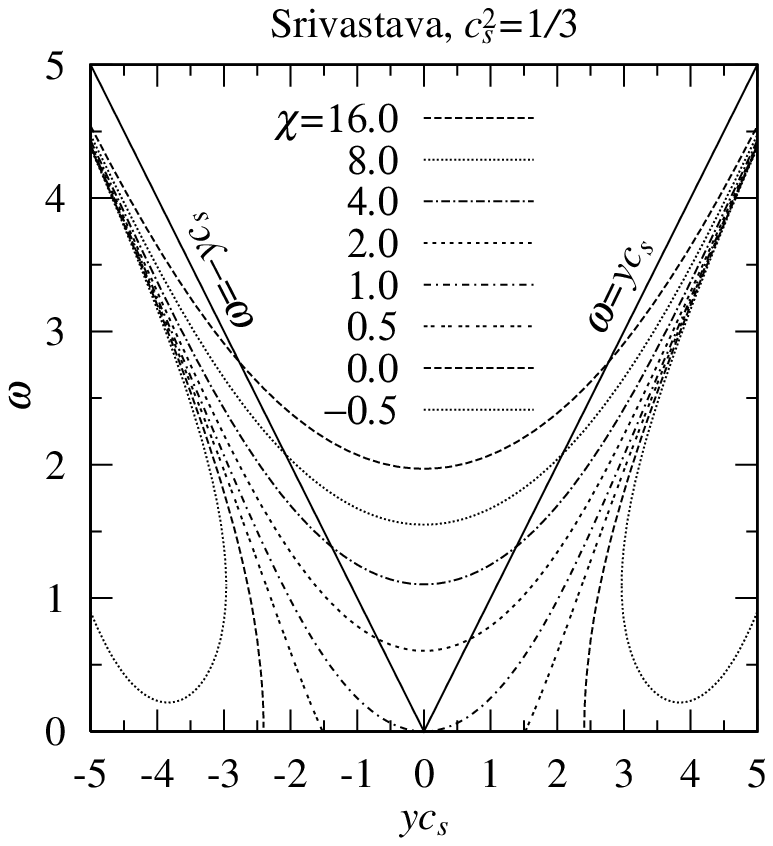}}\\
     (e)\resizebox{0.35\textwidth}{!}{\includegraphics{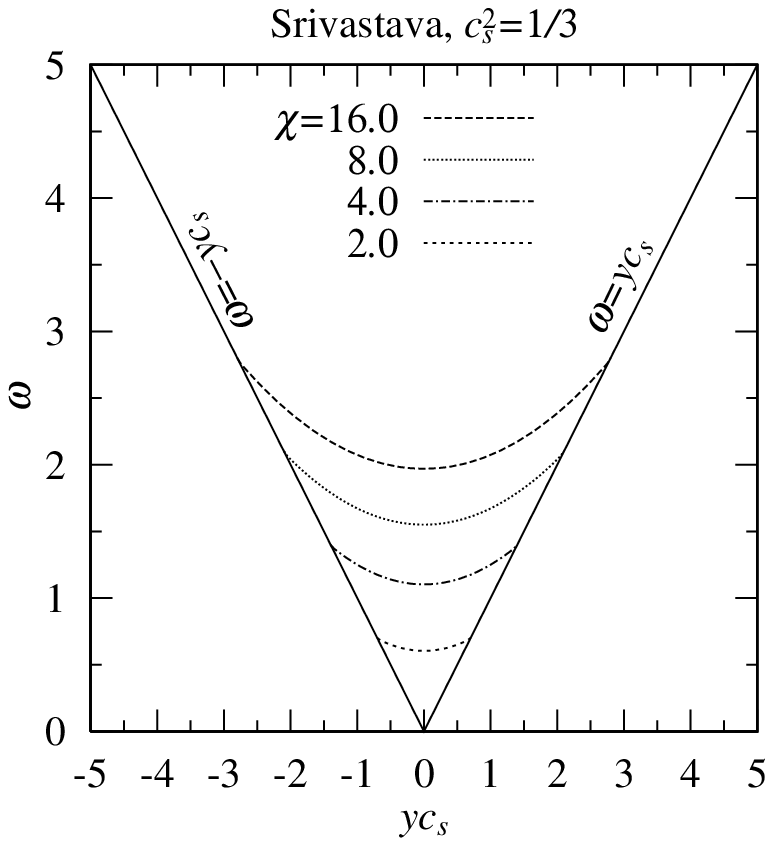}}\\
     
  \end{center}
  \caption{Distributions and contour maps of $\chi$ of eqs.~(\ref{eq_10}) and (\ref{eq_16}). (a) Distribution of eq.~(\ref{eq_10}) with $c_s^2 = 1/3$. (b) Contour map of fig.~\ref{fig_01}(a). (c) Distribution of eq.~(\ref{eq_16}) with $c_s^2 = 1/3$. (d) Contour map of fig.~\ref{fig_01}(c). (e) When assumption with $\chi (\omega,\: y) = 0$ for $\omega < c_s|y|$ is used, this figure is obtained.}
  \label{fig_01}
\end{figure}

On the other hand, Srivastava {\it et al.} have assumed that  $\chi = \chi_1 \exp(\beta \omega)$ and obtained the following equation
\begin{eqnarray}
\diff{^2\chi_1}{y^2} = c_s^2\diff{^2\chi_1}{\omega^2} - c_s^2\beta^2\chi_1 .
\label{eq_13}
\end{eqnarray}
Using a new variable with $p = \beta\sqrt{\omega^2-c_s^2y^2}$,
\begin{eqnarray}
q = \tanh^{-1} \left(\frac{c_sy}{\omega}\right)
\label{eq_14}
\end{eqnarray}
they obtain the following equation which is satisfied with the modified Bessel
function in a special case (as below),
\begin{eqnarray}
\frac 1{p^2}\diff{^2\chi_1}{q^2} - \diff{^2\chi_1}{p^2} 
- \frac 1{p}\diff{\chi_1}{p} + \chi_1 = 0.
\label{eq_15}
\end{eqnarray}
Adopting the condition $\diff{\chi_1}{q}=0$ and assuming the finiteness on the
characteristic ($\omega = \pm c_sy$), they have obtained the following solution,
\begin{eqnarray}
\chi \sim \left\{
\begin{array}{l}
e^{\beta\omega}I_0(\beta \sqrt{\omega^2 - c_s^2y^2})\quad 
{\rm for }\ \omega^2 > c_s^2y^2 ,\smallskip\\
e^{\beta\omega}J_0(\beta \sqrt{c_s^2y^2 - \omega^2})\quad 
{\rm for }\ \omega^2 < c_s^2y^2 .
\end{array}
\right.
\label{eq_16}
\end{eqnarray}
To obtain the potential with the Bessel function, they have introduced a different variable, $q'=\tanh^{-1}(\omega/c_sy)$, in their calculation. To show some mathematical properties of two solutions, we present their contour maps in fig.~\ref{fig_01}. 

The entropy distribution related to eq.~(\ref{eq_16}) is calculated as
\begin{eqnarray}
\frac{dS}{dy} & \sim & \left[ \beta (\beta+1) I_0(p)
+ \frac{\beta^2 ( p^2 + p^2 (2\beta+1) \omega-\beta^2\omega^2)I_0'(p) }{p^3} 
+ \frac{\beta^4 \omega^2 I_0''(p)}{p^2} \right] 
\label{eq_17}
\end{eqnarray}
for $\omega^2 > c_s^2y^2$. For $\omega^2 < c_s^2y^2$ we have
\begin{eqnarray}
\frac{dS}{dy} & \sim & \left[ \beta (\beta+1) J_0(\tilde{p})
+ \frac{\beta^2 ( \tilde{p}^2 + \tilde{p}^2 (2\beta+1) \omega-\beta^2\omega^2)J_0'(\tilde{p}) }{\tilde{p}^3} 
+ \frac{\beta^4 \omega^2 J_0''(\tilde{p})}{\tilde{p}^2} \right] 
\label{eq_18}
\end{eqnarray}
where $\tilde{p} = \beta \sqrt{c_s^2 y^2 - \omega^2}$. Authors of refs.~\cite{Srivastava:1992xb,Mohanty:2003va,Shao:2006sw} have applied eqs.~(\ref{eq_12}), (\ref{eq_17}) and (\ref{eq_18}) to analyses of data. 


\section{Solution including the Heaviside function}
\label{solution2}
Hereafter, to consider a new solution, we use the method of Green functions~\cite{Duffy,Stakgold,Namiki}
\begin{eqnarray}
\diff{^2\chi}{\omega^2} + \left(\frac 1{c_s^2}-1\right)\diff{\chi}{\omega} 
- \frac 1{c_s^2}\diff{^2\chi}{y^2} 
&=& \frac 1{c_s^2}\delta (\omega -\omega_0)\delta (y-y_0).
\quad 
(\omega_0,\ y_0 \to 0)
\label{eq_19}
\end{eqnarray}
Using the function $\chi = \chi_1 e^{\beta \omega}$, we choose the following initial conditions in the ($\omega, y$) space, because we can consider the initial conditions at $\omega = 0$,
\begin{eqnarray}
\left. \diff{\chi_1}{\omega}\right|_{\omega=0+} \!\!\!\!\!\! = G(y) = \delta (y),\quad  \chi_1 (\omega=0+,\: y) = g(y) = 0 .
\label{eq_20}
\end{eqnarray}
The reason is as follows: In the localized point ($(\omega,\: y)\approx (0,\: 0)$), the potential should exist as the boundary value (as the initial condition), then we choose $\delta (y)$, because the fluid created in collisions is localized there. Moreover, the initial condition of the potential at $\omega = 0$ is chosen as $\chi (\omega = 0+,\, y) = 0$. See eq.~(\ref{eq_09}). This set of initial conditions suggests us that the present approach mainly describes the collision in the central region.

\begin{figure}[htbp]
  \begin{center}
    (a)\resizebox{0.45\textwidth}{!}{\includegraphics{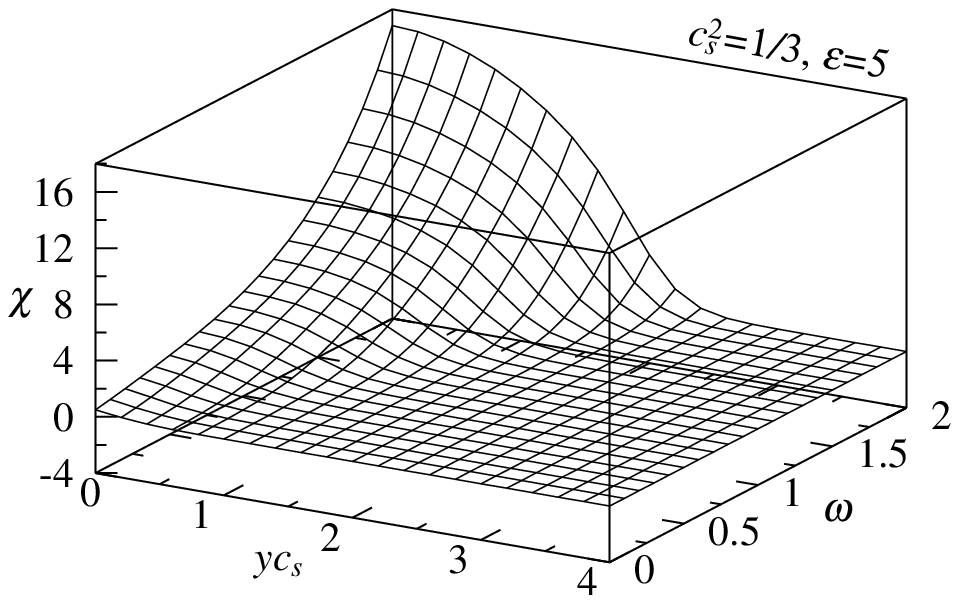}}
    (b)\resizebox{0.35\textwidth}{!}{\includegraphics{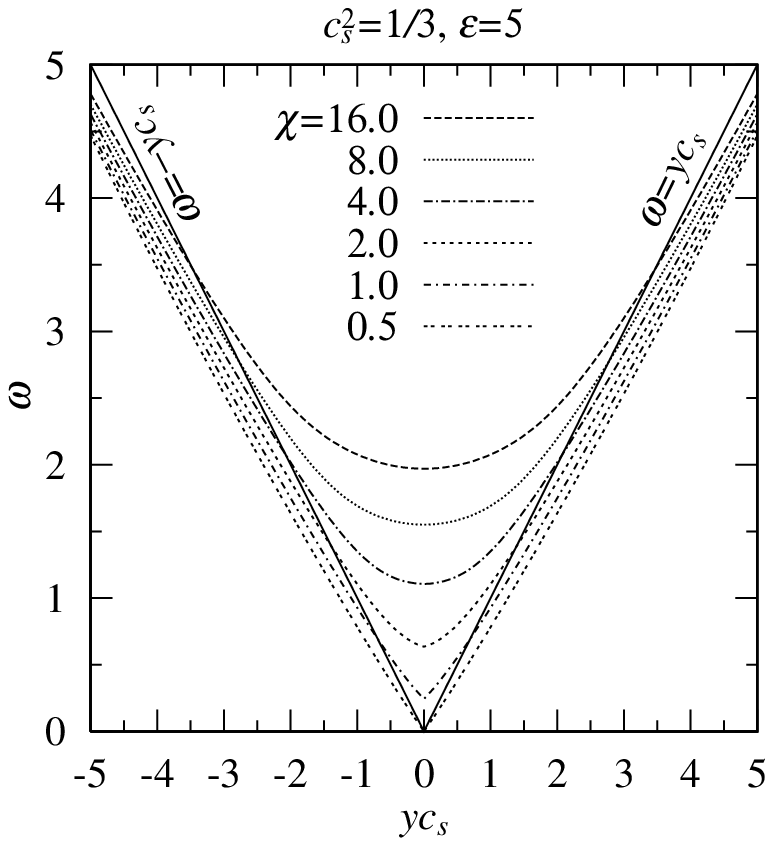}}\\
    (c)\resizebox{0.35\textwidth}{!}{\includegraphics{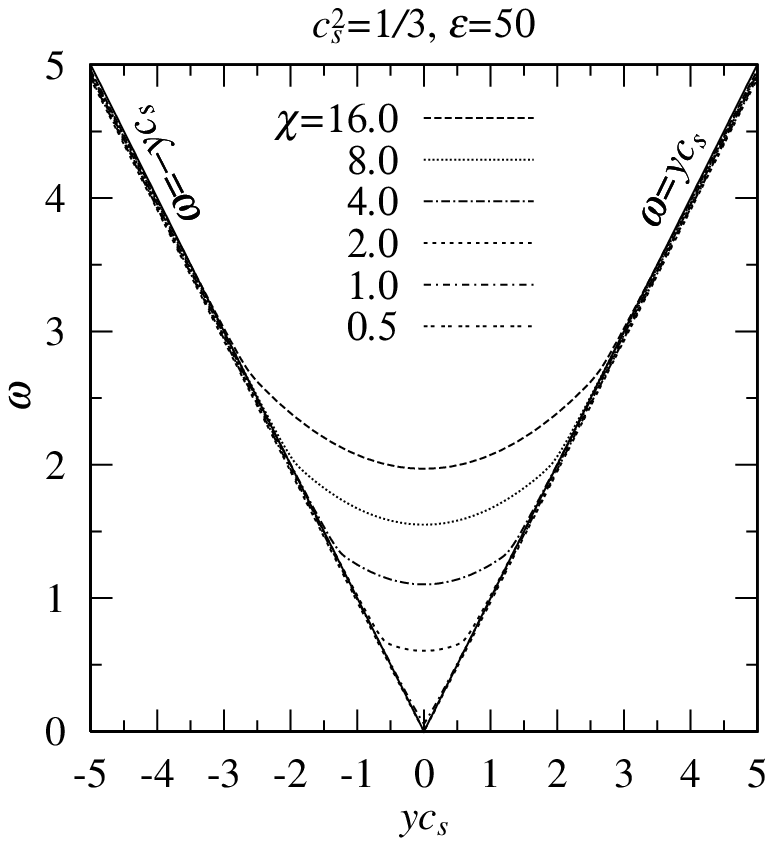}}
  \end{center}
  \caption{(a) (b) Distribution and contour map of eq.~(\ref{eq_22}) with $\varepsilon = 5$. Behaviors distinguished from that of fig.~\ref{fig_01} are observed along the lines $\omega = \pm yc_s$. (c) Contour map of eq.~(\ref{eq_22}) with $\varepsilon = 50$. The figure is the same as fig.~\ref{fig_01}(e) as $\varepsilon = \infty$.}
  \label{fig_02}
\end{figure}

\begin{figure}[htbp]
  \begin{center}
    (a)\resizebox{0.35\textwidth}{!}{\includegraphics{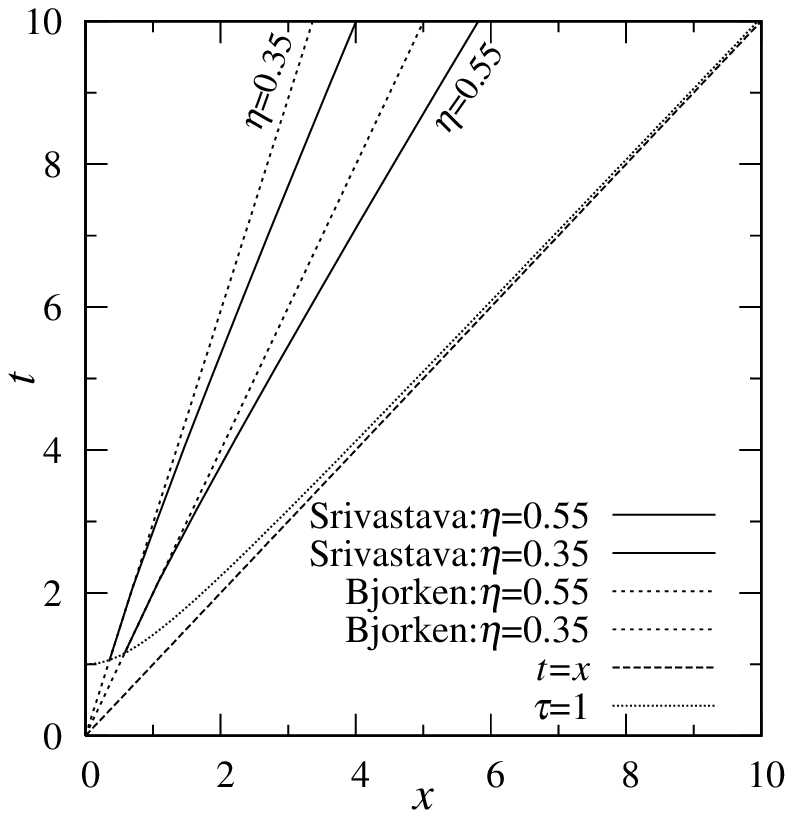}}
    (b)\resizebox{0.35\textwidth}{!}{\includegraphics{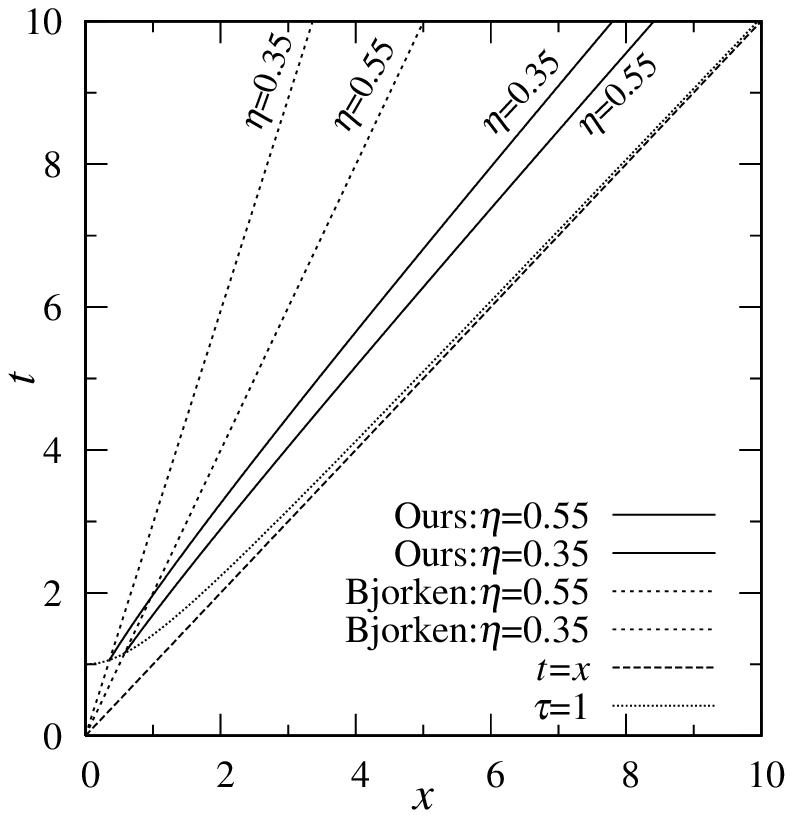}}
  \end{center}
  \caption{Trajectries of $t=\sqrt{\tau^2+x^2}$ ($\tau = 1$ fm). (a) Boost invariant solution by Hwa and Bjorken (dashed lines) and boost non-invariant one by Srivastava {\it et al.} (solid line). (b) Boost invariant solution (dashed lines) and our boost non-invariant one by us with $\varepsilon = 2.5$ (solid line). Finite difference method is used.}
  \label{fig_03}
\end{figure}

In order to obtain the potential governing the hydro-system in collisions, we perform the following calculation named the Riemann's formula,
\begin{eqnarray}
\chi_1 (y,\omega ) & = & \frac{1}{2} \{ g(y+\omega /c_s) 
+ g(y-\omega /c_s) \} \nonumber \\
&+& \frac{1}{2}\int^{\omega /c_s}_{-\omega /c_s} 
\Biggl\{c_s G(z+y) I_0( \beta \sqrt{\omega ^2 - c_s^2z^2}) 
 +  \beta ^2 c_s\omega g(z+y)\frac{I_1(\beta \sqrt{\omega ^2 - c_s^2 z^2})}{\beta \sqrt{\omega ^2 - c_s^2 z^2}}\Biggr\} dz .
\label{eq_21}
\end{eqnarray}
Using eq.~(\ref{eq_20}), we have the following solution
\begin{eqnarray}
\chi \sim e^{\beta\omega}I_0\left(\beta\sqrt{\omega^2-c_s^2y^2}\right)\,H(Q=\omega - c_s|y|) ,
\label{eq_22}
\end{eqnarray}
where $H(Q)$ is the Heaviside function defined as follows
\begin{eqnarray}
H(Q=\omega - c_s|y|) =\left\{
\begin{array}{l}
1\quad Q > 0\ ,\\
0\quad Q < 0\ .
\end{array}
\right.
\nonumber
\end{eqnarray}
The Heaviside function in eq.~(\ref{eq_22}) appears from the region of integration in eq.~(\ref{eq_21}). Hereafter, in concrete analyses, introducing a parameter $\varepsilon$, we use the following approximate expression,
\begin{eqnarray}
H(\varepsilon,\: Q) = \frac 1{e^{-\varepsilon Q} + 1}\ .
\label{eq_23}
\end{eqnarray}
To show some mathematical properties of the above solution, we present its contour map in fig.~\ref{fig_02}. Moreover, we show diagrams between $(t,\ x)$ for eq.~(\ref{eq_16}) and (\ref{eq_22}) in fig.~\ref{fig_03} (a), (b). In fig.~\ref{fig_03} (a) we confirm calculations in ref.~\cite{Srivastava:1992cg}, and in fig.~\ref{fig_03} (b) our calculation for eq.~(\ref{eq_22}) is presented.

\begin{figure}
  \begin{center}
    (a)\resizebox{0.35\textwidth}{!}{\includegraphics{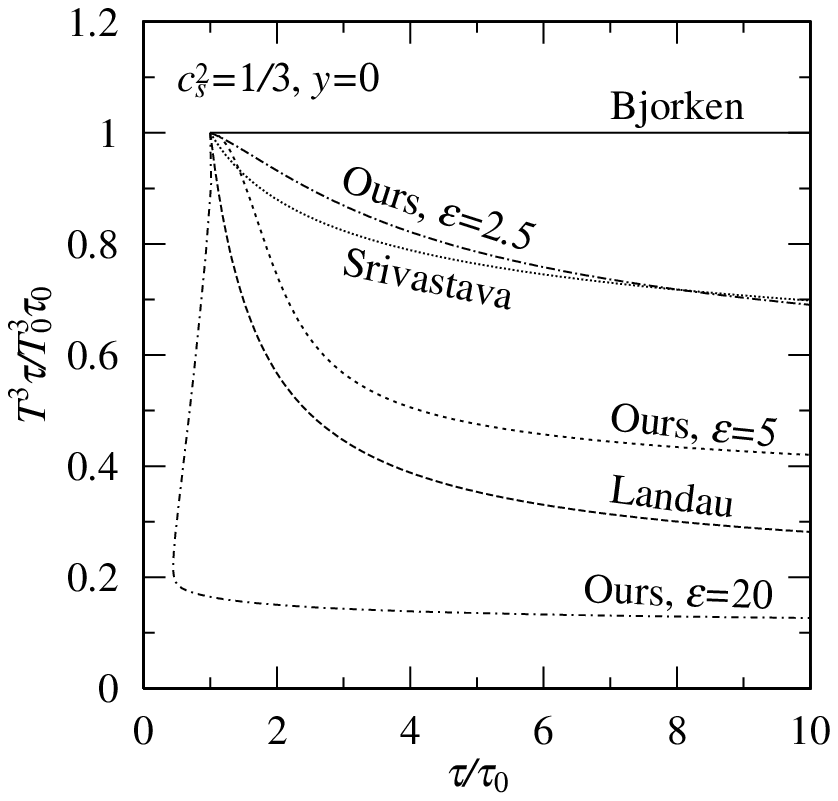}}
    (b)\resizebox{0.35\textwidth}{!}{\includegraphics{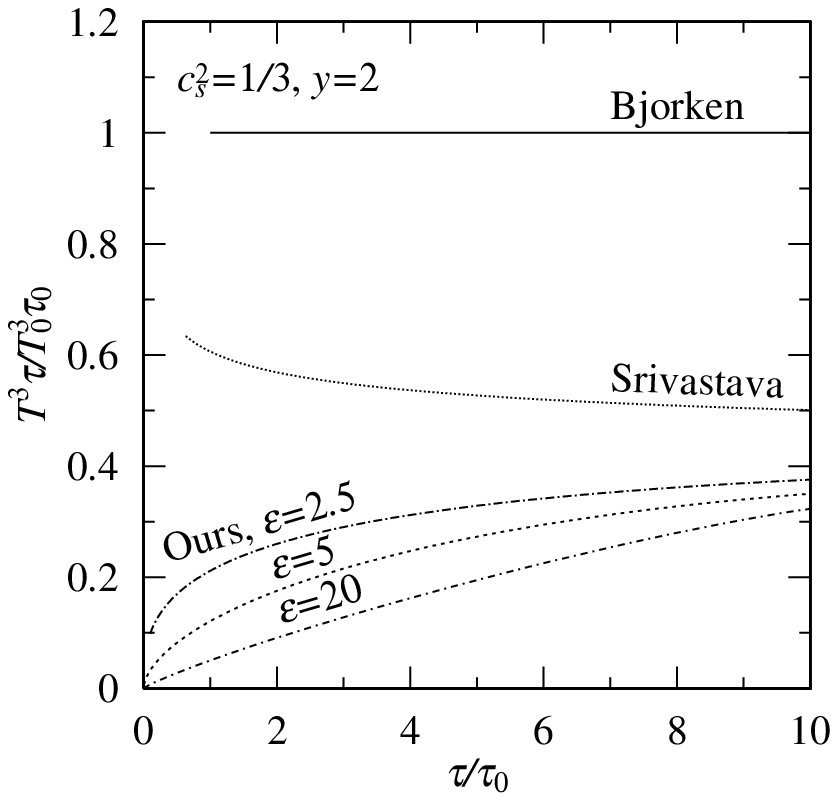}}
  \end{center}
  \caption{The temperatures space-time evolution at $y=0$ and $y=2$ by means of eq.~(\ref{eq_25}) are displayed. Notice that results at $y=0$ by Srivastava {\it et al.} and ours with small $\varepsilon = 2.5$ are the almost same. Ours with $\varepsilon = 5$ is similar to result by Landau. At $y=2$, we cannot estimate a figure by Landau's original solution, because of constraints in integrations.}
  \label{fig_04}
\end{figure}

Here we examine an evolution of the temperature as, 
\begin{eqnarray}
\frac{\tau}{\tau_0}
 = \sqrt{\frac{\frac{1}{T^2}\left[\left(\diff{\chi}{\omega}\right)^2 - \left(\diff{\chi}{y}\right)^2\right]}
 {\frac{1}{T_0^2}\left[\left(\diff{\chi}{\omega}\right)^2 - \left(\diff{\chi}{y}\right)^2\right]_{\omega=0,\: y=0}}}
 = \sqrt{\frac{e^{2\omega}\left[\left(\diff{\chi}{\omega}\right)^2 - \left(\diff{\chi}{y}\right)^2\right]}
 {\left[\left(\diff{\chi}{\omega}\right)^2 - \left(\diff{\chi}{y}\right)^2\right]_{\omega=0,\: y=0}}}
\label{eq_24}
\end{eqnarray}
and
\begin{eqnarray}
\frac{T^3}{T_0^3}\frac{\tau}{\tau_0} = e^{-3\omega}\frac{\tau}{\tau_0}
\label{eq_25}
\end{eqnarray}
The results at $y=0$ and $y=2$ are presented in fig.~\ref{fig_04}. We can compare differences of the temperatures evolution among refs.~\cite{Srivastava:1992cg,Landau:1953gs,Bjorken:1982qr} and ours.

\begin{figure}
  \begin{center}
    (a)\resizebox{0.35\textwidth}{!}{\includegraphics{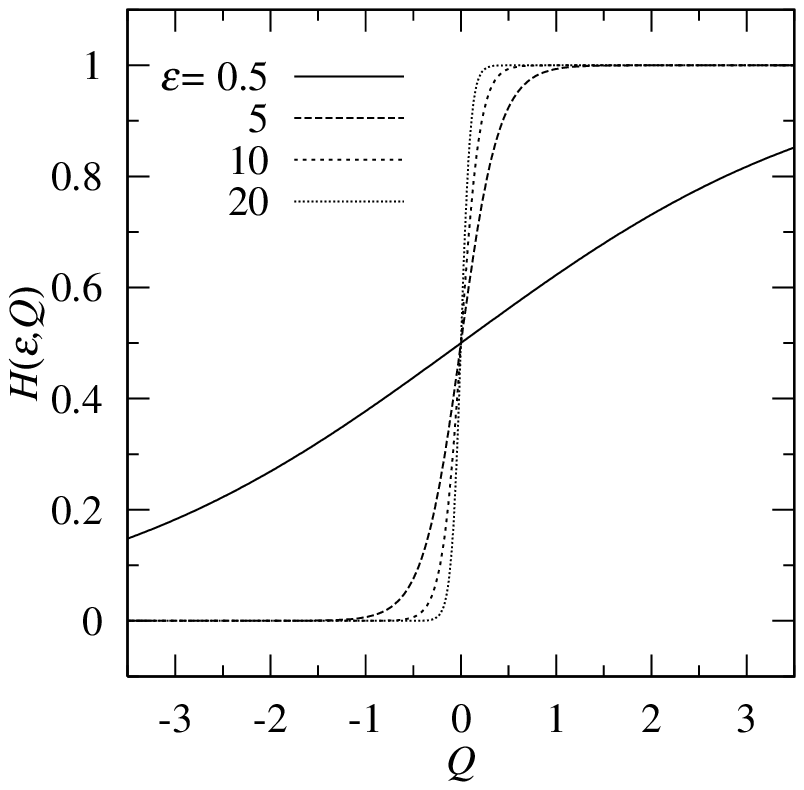}}
    (b)\resizebox{0.35\textwidth}{!}{\includegraphics{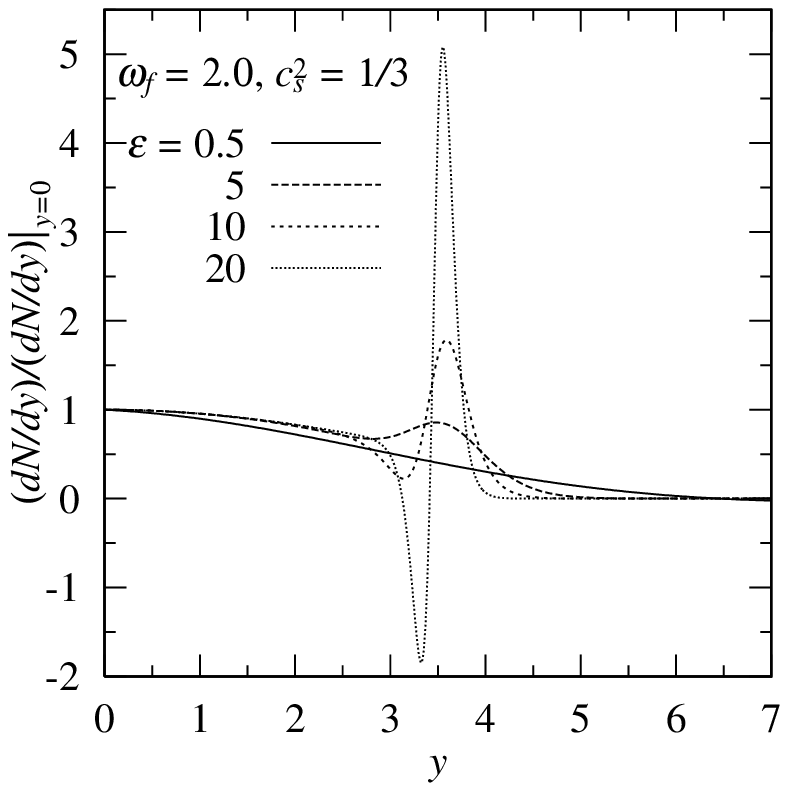}}
  \end{center}
  \caption{(a) Behavior of eq.~(\ref{eq_23}). (b) Behavior of eq.~(\ref{eq_26}).}
  \label{fig_05}
\end{figure}

The following expression is the entropy distribution,
\begin{eqnarray}
\frac{dN}{dy} \sim \frac{dS}{dy} & \sim & \biggl[ \beta (\beta+1) I_0(p) 
 + \frac{\beta^2 ( p^2 + p^2 (2\beta+1) \omega-\beta^2\omega^2) I_0'(p)}{p^3} 
 + \frac{\beta^4 \omega^2 I_0''(p)}{p^2}\biggr] H(\varepsilon,\ Q)\nonumber\\
  && + \biggl[ (2\beta+1)I_0(p)
  + \frac{2\beta^2 \omega I_0'(p)}p \biggr] H'(\varepsilon,\ Q)
 + I_0(p) H''(\varepsilon,\ Q) .
\label{eq_26}
\end{eqnarray}
When $p$ is imaginary, the modified Bessel function should be replaced by the Bessel functions, because we treat $\varepsilon$ as a free parameter in the CERN MINUIT program. This means that we are able to obtain any useful information for collisions, provided that there was a relation among $\varepsilon$, $\omega$ and $c_s$ governed collisions. The typical behavior of eq.~(\ref{eq_26}) is shown in fig.~\ref{fig_05}. 

To analyze the measured distribution of hadrons, the following calculation is necessary,
\begin{eqnarray}
\frac{dN}{dy_h} = c \int_{-y_0}^{y_0} dy\int_0^{\infty} dp_T\, p_T 
 \frac{m_T\cosh(y_h-y)}{\exp[\{ m_T\cosh(y_h-y) -\mu_B \}/T] + \delta}\frac{dN}{dy},
\label{eq_27}
\end{eqnarray}
where $y_h$, $p_T$,$m_T = \sqrt{p_T^2+ m_p^2}$ and $\mu_A$ are the rapidity of hadron, the transverse momentum of hadron and the baryonic chemical potential of hadron A, respectively. For mesons, $\delta = -1,~\mu_B = 0$ and for baryons $\delta = 1,~\mu_B = \mu$. As you see fig.~\ref{fig_05}, due to the derivative of the Heaviside function, there are small-finite contributions outside of the asymptotic lines $\omega = \pm c_s|y|$. 


\section{Analyses of data by several formulae}
\label{analyses}
\begin{figure}[htbp]
  \begin{center}
    \resizebox{0.41\textwidth}{!}{\includegraphics{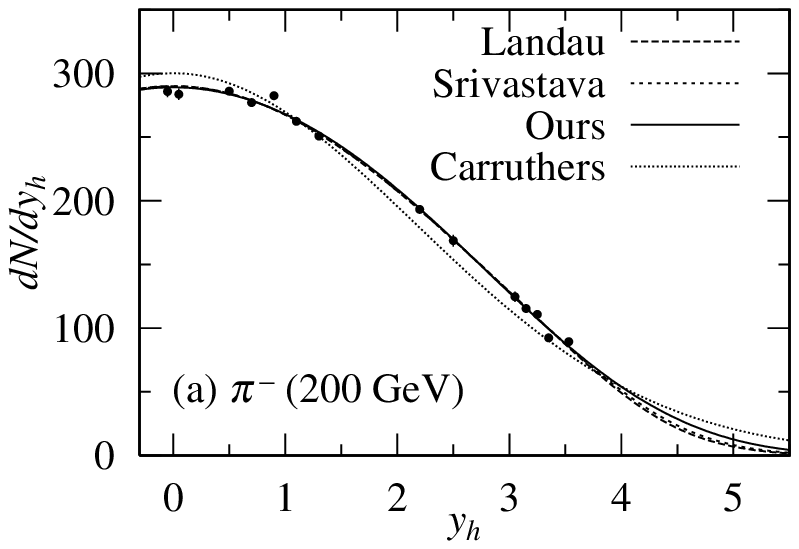}}
    \resizebox{0.41\textwidth}{!}{\includegraphics{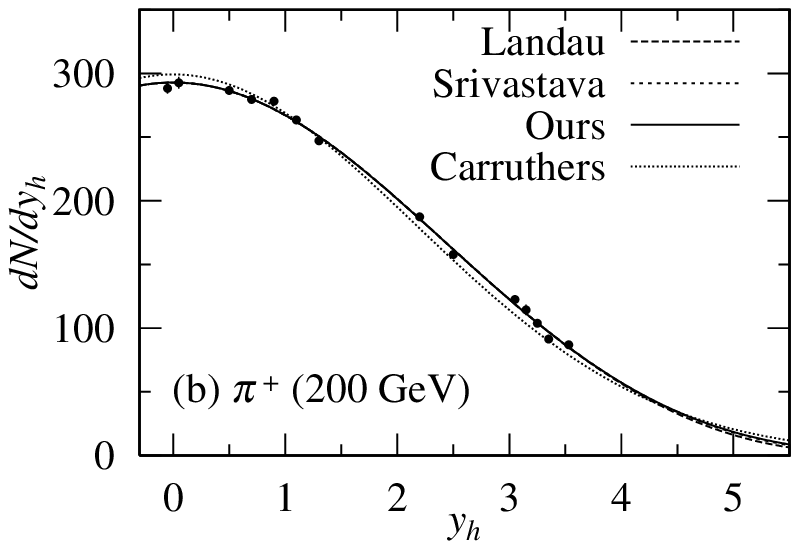}}\\
    \resizebox{0.41\textwidth}{!}{\includegraphics{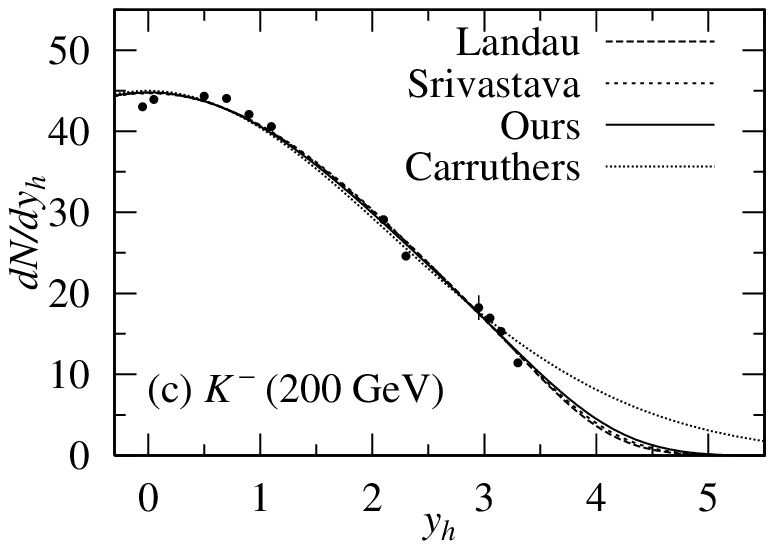}}
    \resizebox{0.41\textwidth}{!}{\includegraphics{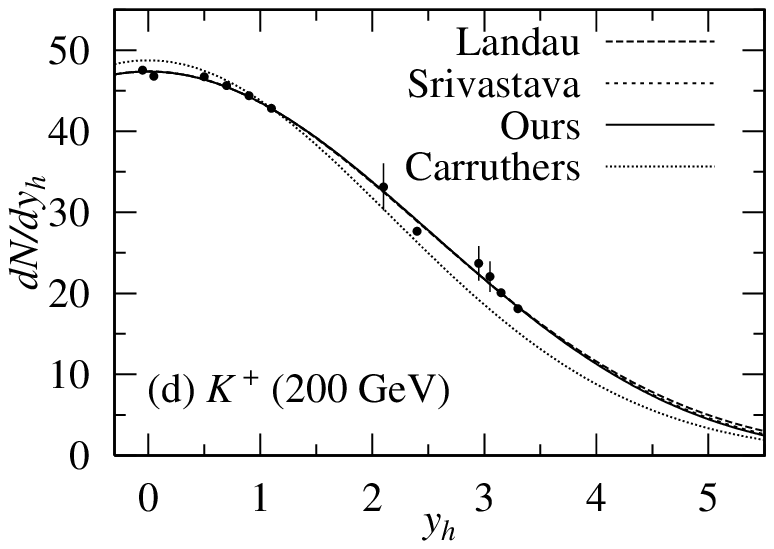}}
  \end{center}
  \caption{Comparison of pions and K mesons rapidity distribution and several formulae at $\sqrt{s_{\tiny NN}} =$ 200 GeV. Results by Srivastava and ours in tables~\ref{table_02} and \ref{table_04} are used.}
  \label{fig_06}
\end{figure}

\begin{figure}[htbp]
  \begin{center}
    \resizebox{0.41\textwidth}{!}{\includegraphics{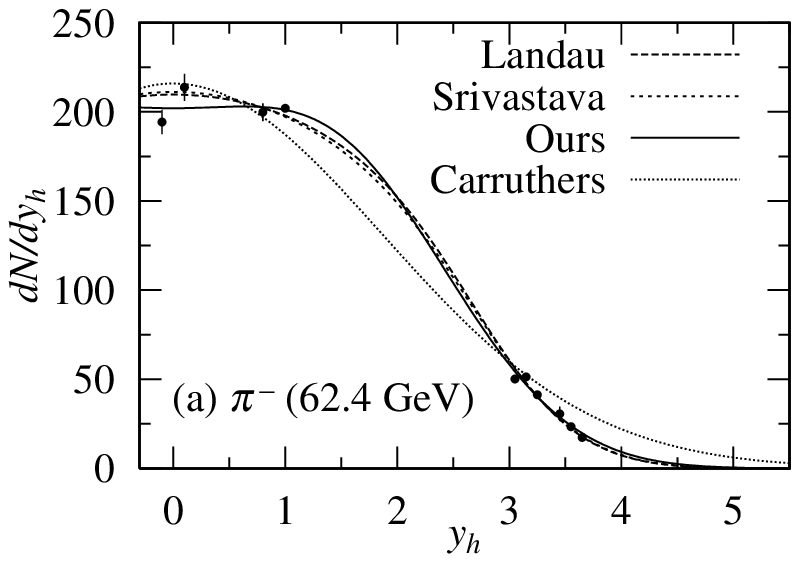}}
    \resizebox{0.41\textwidth}{!}{\includegraphics{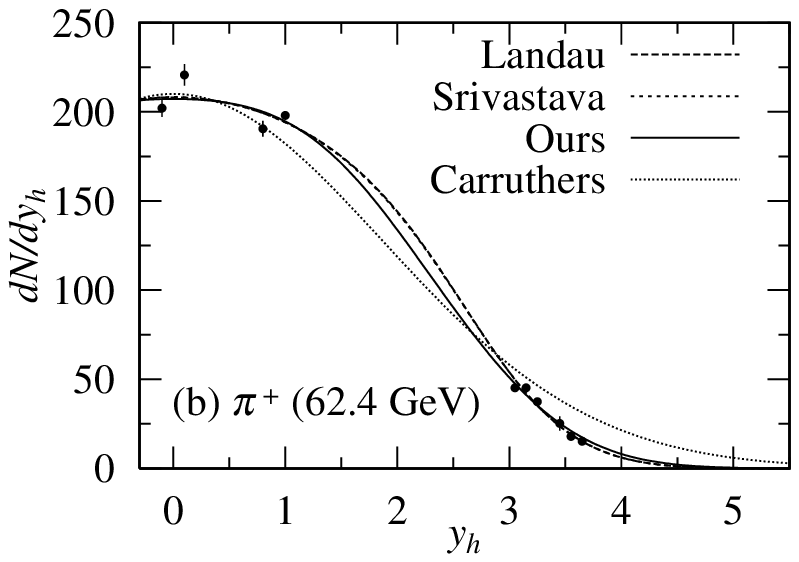}}\\
    \resizebox{0.41\textwidth}{!}{\includegraphics{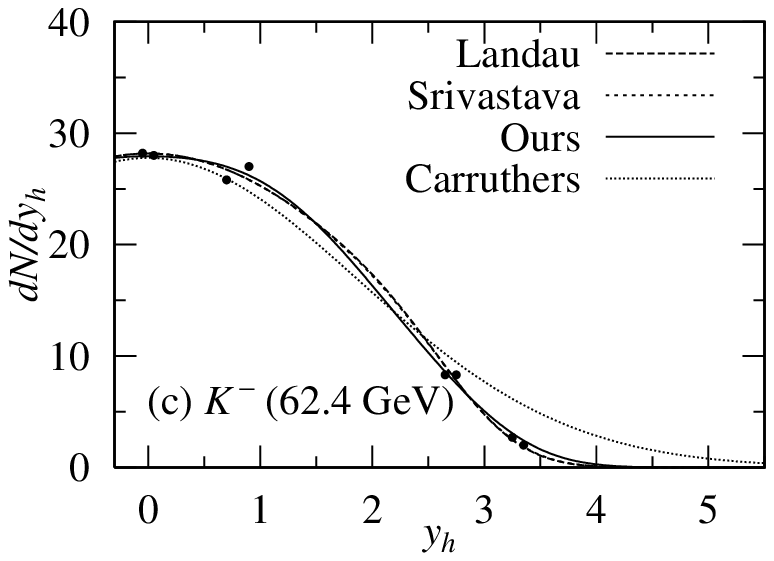}}
    \resizebox{0.41\textwidth}{!}{\includegraphics{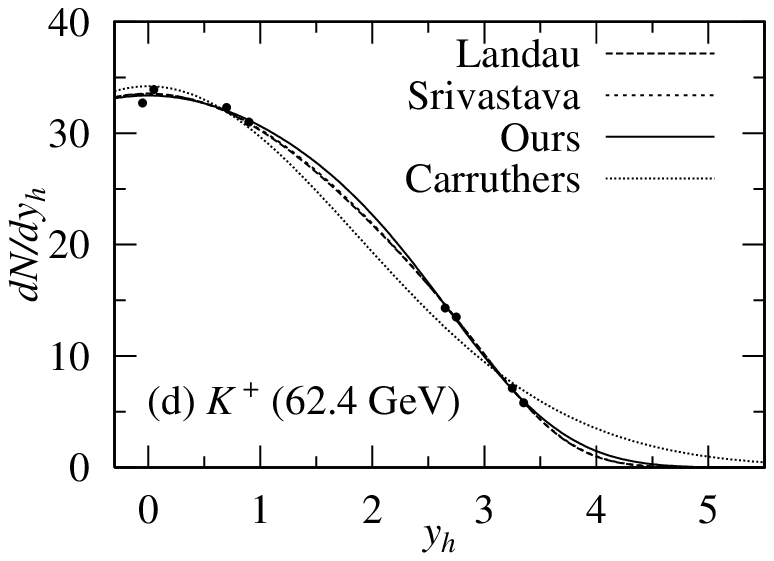}}
  \end{center}
  \caption{Comparison of pions and K mesons rapidity distribution and several formulae at $\sqrt{s_{\tiny NN}} =$ 62.4 GeV. Results by Srivastava and ours in tables~\ref{table_02} and \ref{table_04} are used.}
  \label{fig_07}
\end{figure}

Since BRAHMS Collaboration has reported the rapidity distributions of charged $\pi$ mesons and K mesons at RHIC $\sqrt{s_{\tiny NN}} =$ 62.4 GeV and 200 GeV, we can analyze them by three formulae mentioned above. The freeze-out temperature and chemical potential are cited from ref.~\cite{Andronic:2005yp}. $T_f = 0.1605$ (200 GeV and 62.4GeV) and $\mu = 0$. See also ref.~\cite{Biyajima:2006mv}. Moreover, we can analyze the data by the Gaussian distribution proposed by Carruthers and Duong-Van~\cite{Carruthers:1973rw}.
\begin{eqnarray}
\frac{dN}{dy_h} = \frac c{\sqrt{2\pi}\sigma}\exp\left[-\frac{y_h^2}{2\sigma^2}\right],\quad \sigma^2 = \ln \left(\frac{\sqrt{s_{\tiny NN}}}{2m_p}\right)
\label{eq_28}
\end{eqnarray}
Our results by eq.~(\ref{eq_27}) in which $c_s^2$ is assumed as a free parameter are presented in figs.~\ref{fig_06} and \ref{fig_07}. Notice that we examine different cutoff values for $y_0$ in calculations of eq.~(\ref{eq_27}). Analyses of data by eq.~(\ref{eq_28}) are cited in table~\ref{table_06} and shown in figs.~\ref{fig_06} and \ref{fig_07}.

\begin{table}[htbp]
\begin{center}
\caption{Analysis of charged pions and K mesons rapidity distributions by means of eqs.~(\ref{eq_12}) and (\ref{eq_27}) with $y_0 = \omega/\sqrt{c_s^2}$. (Distribution by Landau)}
\label{table_01}
\vspace{2mm}
\begin{tabular}{ccccc}
\hline
 data  & $\omega$  &  $c_s^2$  & $c$  &  $\chi^2$/n.d.f \\
\hline
$\pi^-$ (200 GeV)  & 1.77$\pm$0.03 & 0.194$\pm$0.016 & 1230$\pm$426      & 18/11 \\
$\pi^+$ (200 GeV)  & 1.84$\pm$0.02 & 0.135$\pm$0.001 & 169$\pm$12        & 10/11 \\
$K^-$ (200 GeV)    & 1.58$\pm$0.05 & 0.179$\pm$0.024 & 691$\pm$371       & 6.3/9 \\
$K^+$ (200 GeV)    & 2.49$\pm$0.02 & 0.066$\pm$0.001 & 0.0013$\pm$0.0002 & 1.8/9 \\
$\pi^-$ (62.4 GeV) & 1.61$\pm$0.01 & 0.333$\pm$0.059 & 4860$\pm$92       & 13/7 \\
$\pi^+$ (62.4 GeV) & 1.49$\pm$0.16 & 0.295$\pm$0.074 & 4250$\pm$1580     & 15/7 \\
$K^-$ (62.4 GeV)   & 1.22$\pm$0.23 & 0.189$\pm$0.107 & 1040$\pm$1250     & 1.3/5 \\
$K^+$ (62.4 GeV)   & 1.40$\pm$0.08 & 0.178$\pm$0.043 & 724$\pm$503       & 0.44/5 \\
\hline
\end{tabular}
\end{center}
\end{table}

\begin{table}[htbp]
\begin{center}
\caption{Analysis of charged pions and K mesons rapidity distributions by means of eqs.~(\ref{eq_17}), (\ref{eq_18}) and (\ref{eq_27}) with $y_0 = \omega/\sqrt{c_s^2} + \delta\ (\delta = 2)$. (Distribution by Srivastava {\it et al.})}
\label{table_02}
\vspace{2mm}
\begin{tabular}{ccccc}
\hline
 data  & $\omega$  &  $c_s^2$  & $c$  &  $\chi^2$/n.d.f \\
\hline
 $\pi^-$ (200 GeV)  & 1.09$\pm$0.15 & 0.246$\pm$0.024 & 1140$\pm$659      & 17/11 \\
 $\pi^+$ (200 GeV)  & 1.73$\pm$0.02 & 0.128$\pm$0.001 & 6.52$\pm$0.46     & 10/11 \\
 $K^-$ (200 GeV)    & 0.90$\pm$0.20 & 0.236$\pm$0.032 & 657$\pm$500       & 6.4/9 \\
 $K^+$ (200 GeV)    & 2.24$\pm$0.04 & 0.102$\pm$0.001 & 0.0524$\pm$0.0090 & 1.8/9 \\
 $\pi^-$ (62.4 GeV) & 0.48$\pm$0.02 & 0.333$\pm$0.033 & 4120$\pm$77       & 14/7 \\
 $\pi^+$ (62.4 GeV) & 0.43$\pm$0.02 & 0.333$\pm$0.245 & 4200$\pm$211      & 15/7 \\
 $K^-$ (62.4 GeV)   & 0.41$\pm$0.13 & 0.244$\pm$0.098 & 839$\pm$1100      & 1.3/5 \\
 $K^+$ (62.4 GeV)   & 0.66$\pm$0.12 & 0.240$\pm$0.039 & 703$\pm$464       & 0.42/5 \\
\hline
\end{tabular}
\end{center}
\end{table}

\begin{table}[htbp]
\begin{center}
\caption{Analysis of charged pions and K mesons rapidity distributions by means of eqs.~(\ref{eq_17}), (\ref{eq_18}) and (\ref{eq_27}) with $y_0 = \ln (\sqrt{s_{\tiny NN}}/m_p)$. (Distribution by Srivastava {\it et al.})}
\label{table_03}
\vspace{2mm}
\begin{tabular}{ccccc}
\hline
 data  &  $\omega$  &  $c_s^2$  & $c$  &  $\chi^2$/n.d.f \\
\hline
 $\pi^-$ (200 GeV)  & 0.87$\pm$0.68               & 0.268$\pm$0.065 & 2030$\pm$2910     & 17/11 \\
 $\pi^+$ (200 GeV)  & 1.73$\pm$0.02               & 0.128$\pm$0.001 & 6.51$\pm$0.46     & 10/11 \\
 $K^-$ (200 GeV)    & 0.69$\pm$1.59               & 0.255$\pm$0.259 & 1100$\pm$5110     & 6.5/9 \\
 $K^+$ (200 GeV)    & 2.24$\pm$0.04               & 0.102$\pm$0.001 & 0.0542$\pm$0.0093 & 1.8/9 \\
 $\pi^-$ (62.4 GeV) & 4.12$\times 10^{-7}\pm$0.35 & 0.265$\pm$0.002 & 3470$\pm$62       & 21/7 \\
 $\pi^+$ (62.4 GeV) & 1.40$\times 10^{-6}\pm$0.60 & 0.259$\pm$0.002 & 3230$\pm$59       & 18/7 \\
 $K^-$ (62.4 GeV)   & 4.55$\times 10^{-5}\pm$9.84 & 0.247$\pm$0.011 & 1330$\pm$209      & 1.4/5 \\
 $K^+$ (62.4 GeV)   & 1.17$\times 10^{-6}\pm$6.47 & 0.283$\pm$0.008 & 2120$\pm$134      & 0.56/5 \\
\hline
\end{tabular}
\end{center}
\end{table}

\begin{table}[htbp]
\begin{center}
\caption{Analysis of charged pions and K mesons rapidity distributions by means of eqs.~(\ref{eq_26}) and (\ref{eq_27}) with $y_0 = \omega/\sqrt{c_s^2} + \delta\ (\delta = 2)$. (Our distribution)}
\label{table_04}
\vspace{2mm}
\begin{tabular}{cccccc}
\hline
 data  &  $\omega$  &  $c_s^2$  &  $c$  &  $\varepsilon$  &  $\chi^2$/n.d.f \\
\hline
 $\pi^-$ (200 GeV)  & 1.55$\pm$0.15 & 0.276$\pm$0.052 & 1120$\pm$1190   & 1.71$\pm$0.44  & 16/10 \\
 $\pi^+$ (200 GeV)  & 1.86$\pm$0.02 & 0.112$\pm$0.001 & 1.35$\pm$0.09   & 3.35$\pm$0.80  & 10/10 \\
 $K^-$ (200 GeV)    & 1.40$\pm$0.14 & 0.165$\pm$0.010 & 50.9$\pm$18.6   & 6.15$\pm$7.66  & 6.6/8 \\
 $K^+$ (200 GeV)    & 2.28$\pm$0.04 & 0.105$\pm$0.001 & 0.064$\pm$0.010 & 2.50$\pm$1.61    & 1.9/8 \\
 $\pi^-$ (62.4 GeV) & 0.86$\pm$0.02 & 0.333$\pm$0.331 & 2870$\pm$80     & 3.22$\pm$0.43  & 8.5/6 \\
 $\pi^+$ (62.4 GeV) & 0.77$\pm$0.03 & 0.333$\pm$0.223 & 3100$\pm$140    &  2.52$\pm$0.33 & 16/6 \\
 $K^-$ (62.4 GeV)   & 0.68$\pm$0.09 & 0.333$\pm$0.267 & 1570$\pm$220    & 2.12$\pm$0.74  & 1.2/4 \\
 $K^+$ (62.4 GeV)   & 1.16$\pm$0.18 & 0.217$\pm$0.132 & 241$\pm$822     & 3.66$\pm$3.86  & 0.40/4 \\
\hline
\end{tabular}
\end{center}
\end{table}

\begin{table}[htbp]
\begin{center}
\caption{Analysis of charged pions and K mesons rapidity distributions by means of eqs.~(\ref{eq_26}) and (\ref{eq_27}) with $y_0 = \ln (\sqrt{s_{\tiny NN}}/m_p)$. (Our distribution)}
\label{table_05}
\vspace{2mm}
\begin{tabular}{cccccc}
\hline
 data  &  $\omega$  &  $c_s^2$  &  $c$  &  $\varepsilon$  &  $\chi^2$/n.d.f \\
\hline
 $\pi^-$ (200 GeV) & 1.55$\pm$0.09 & 0.277$\pm$0.030 & 1150$\pm$690 & 1.70$\pm$0.32 & 16/10 \\
 $\pi^+$ (200 GeV)  & 1.86$\pm$0.02 & 0.113$\pm$0.001 & 1.35$\pm$0.09     & 3.35$\pm$0.80 & 10/10 \\
 $K^-$ (200 GeV)   & 1.37$\pm$0.16 & 0.167$\pm$0.007 &  58.9$\pm$28.3 &  7.31$\pm$6.82 &  6.5/8 \\
 $K^+$ (200 GeV)    & 2.28$\pm$0.04 & 0.105$\pm$0.001 & 0.0639$\pm$0.0098 & 2.50$\pm$1.61 & 1.9/8 \\
 $\pi^-$ (62.4 GeV) & 0.90$\pm$0.08 & 0.306$\pm$0.047 & 2150$\pm$1210     & 3.86$\pm$1.35 & 8.6/6 \\
 $\pi^+$ (62.4 GeV) & 0.93$\pm$0.02 & 0.227$\pm$0.026 & 741$\pm$334       & 6.31$\pm$3.12 & 14/6 \\
 $K^-$ (62.4 GeV)   & 0.89$\pm$0.10 & 0.215$\pm$0.076 & 291$\pm$439       & 4.95$\pm$3.73 & 1.1/4 \\
 $K^+$ (62.4 GeV)   & 1.16$\pm$0.18 & 0.217$\pm$0.246 & 243$\pm$857       & 3.65$\pm$3.93 & 0.40/4 \\
\hline
\end{tabular}
\end{center}
\end{table}

\begin{table}[htbp]
\begin{center}
\caption{Analysis of charged pions and K mesons rapidity distributions by means of eq.~(\ref{eq_28}). (Distribution by Carruthers {\it et al.})}
\label{table_06}
\vspace{2mm}
\begin{tabular}{cccc}
\hline
 data               & $\sigma$ & $c$     & $\chi^2$/n.d.f\\
\hline
 $\pi^-$ (200 GeV)  & 2.16 & 1626$\pm$8  & 101/13 \\
 $\pi^+$ (200 GeV)  & 2.16 & 1621$\pm$8  &  45/13 \\
 $K^-$ (200 GeV)    & 2.16 &  244$\pm$3  & 9.2/11 \\
 $K^+$ (200 GeV)    & 2.16 &  264$\pm$3  &  24/11 \\
 $\pi^-$ (62.4 GeV) & 1.87 & 1013$\pm$11 &  82/9 \\
 $\pi^+$ (62.4 GeV) & 1.87 &  986$\pm$10 & 117/9 \\
 $K^-$ (62.4 GeV)   & 1.87 &  130$\pm$4  &  14/7 \\
 $K^+$ (62.4 GeV)   & 1.87 &  161$\pm$4  & 4.2/7 \\
\hline
\end{tabular}
\end{center}
\end{table}

\section{Concluding remarks}
\label{concluding}
\begin{figure}[htbp]
  \begin{center}
    (a)\resizebox{0.45\textwidth}{!}{\includegraphics{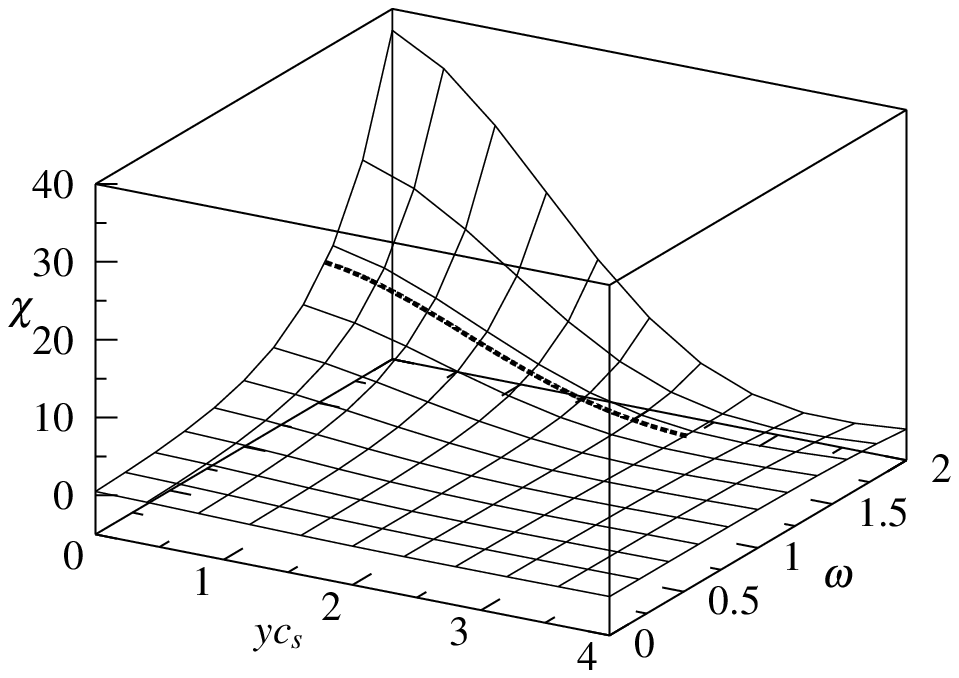}}
    (b)\resizebox{0.45\textwidth}{!}{\includegraphics{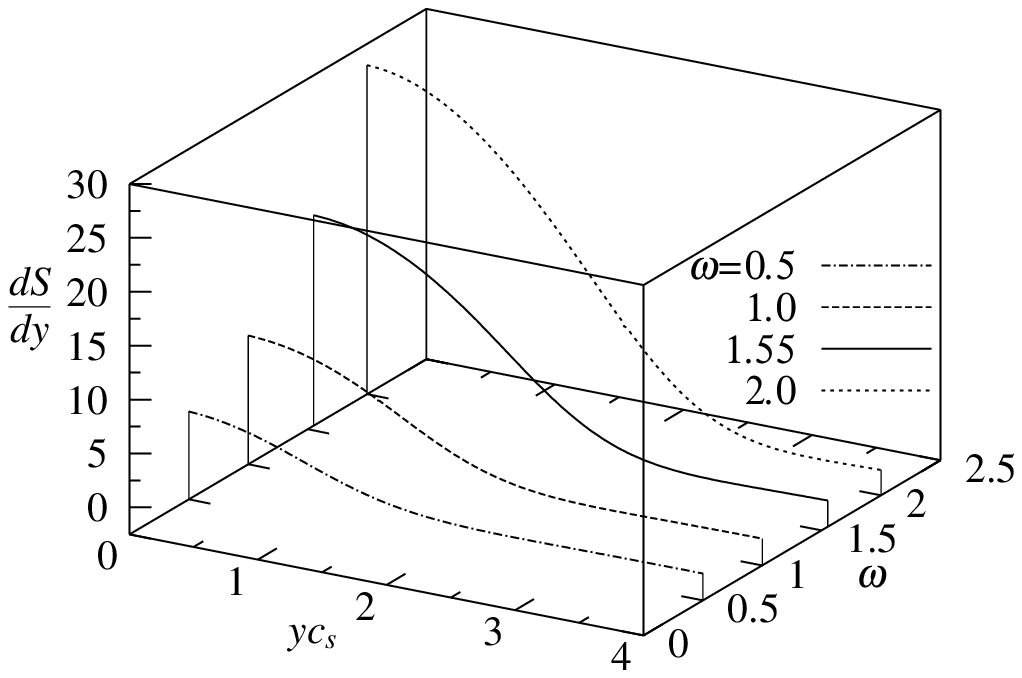}}
  \end{center}
  \caption{(a) Potential of fluid (dashed line) changes on the line of $\omega_f = 1.55$. $c_s^2 = 0.277$ and $\varepsilon = 1.7$ (200 GeV, $\pi^-$). (b) Entropy distribution of fluid.}
  \label{fig_08}
\end{figure}

\begin{figure}[htbp]
  \begin{center}
    \resizebox{0.41\textwidth}{!}{\includegraphics{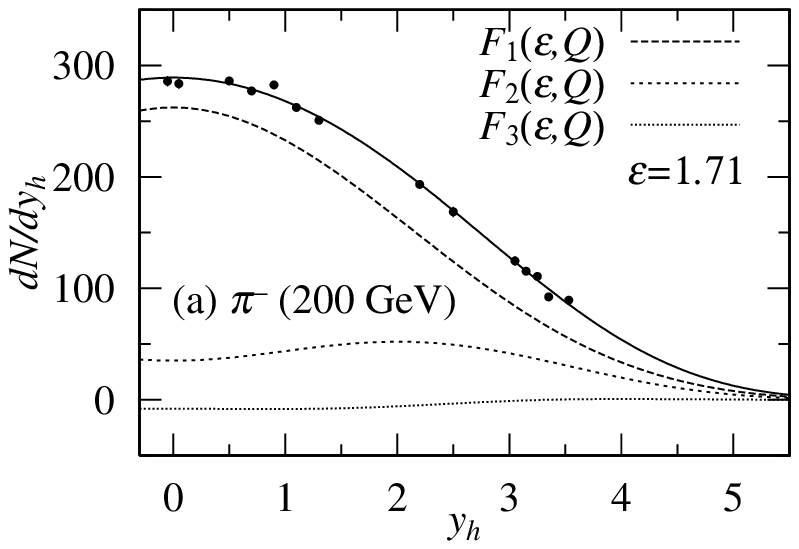}}
    \resizebox{0.41\textwidth}{!}{\includegraphics{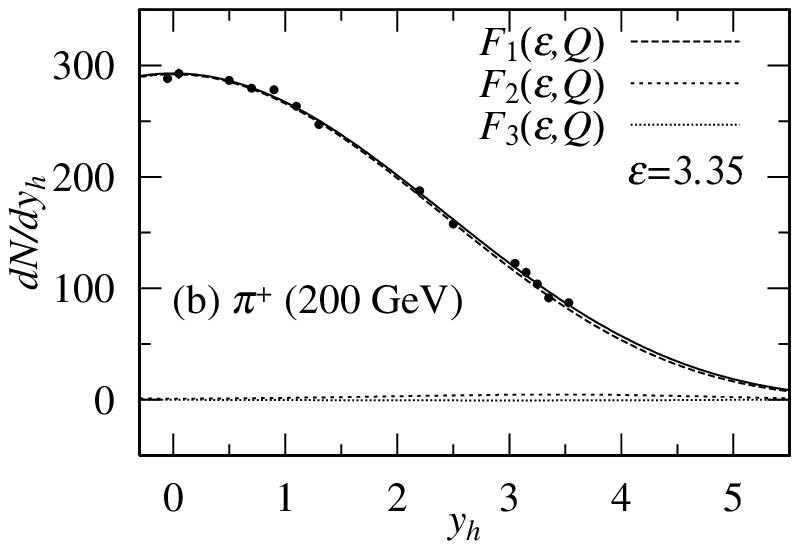}}\\
    \resizebox{0.41\textwidth}{!}{\includegraphics{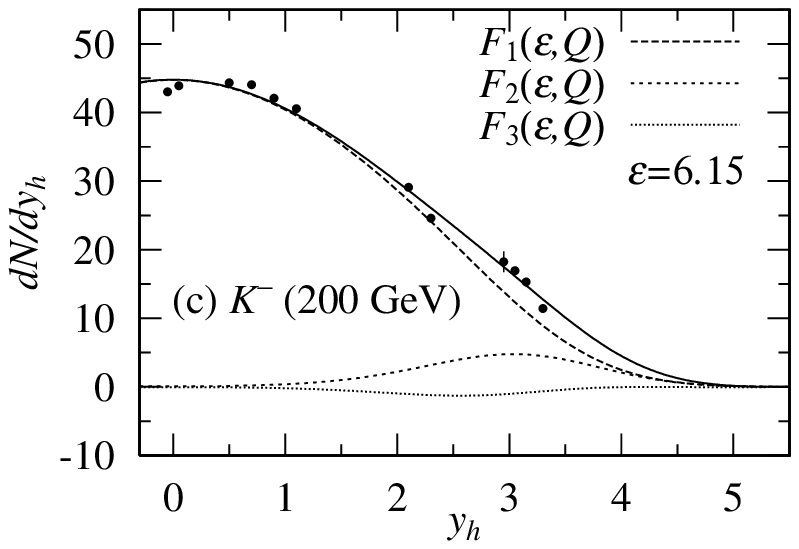}}
  \end{center}
  \caption{Three components of our distributions on charged pions and K mesons at $\sqrt{s_{\tiny NN}} =$ 200 GeV. The first term, second one, and third one of eq.~(\ref{eq_26}) are named $F_1(\varepsilon,\ Q)$, $F_2(\varepsilon,\ Q)$, $F_3(\varepsilon,\ Q)$, respectively. Results in table~\ref{table_04} are used.}
  \label{fig_09}
\end{figure}

We summarize the present studies as follows:\smallskip\\
1) We have investigated the new solution including the Heaviside function in 1+1 dimensional hydrodynamics. The Heaviside function is approximately expressed by eq.~(\ref{eq_23}). Its typical behaviors are shown in fig.~\ref{fig_05}. The concrete potential and the entropy distribution are presented in fig.~\ref{fig_08}. Moreover, three concrete components of our distribution are shown in fig.~\ref{fig_09}. Of course, the second and third ones depend on the values of $\varepsilon$'s. The role of the parameter ƒÃ seems to be reflecting the temperature fluctuation on the characteristic ($\omega = \pm c_sy$) in the present analyses.\smallskip\\
2) Three distributions by Landau, Srivastava {\it et al.}, and ours show almost the same minimum chi-squared as CERN MINUIT is used. However, values of $\omega$'s are different, provided that eq.~(\ref{eq_26}) are applied to the data. Eq.~(\ref{eq_27}) by Srivastava {\it et al.} strongly depends on cutoffs values $y_0$'s at 62.4 GeV. In other words, for values of cutoff no principle exists. The main reason is attributed to their assumption of the finiteness on the characteristic ($\omega = \pm c_sy$). Please compare fig.~\ref{fig_01}(e) and fig.~\ref{fig_02}(c).\smallskip\\
3) On the other hand, our distribution function, due to the Heaviside function, does not strongly depend on cutoffs at both energies. It is interesting that three sound velocities $c_s$'s at 62.4 GeV are 0.333, and decrease at 200 GeV in table~\ref{table_04}. Among the Gaussian distribution and distributions based on 1+1 dimensional hydrodynamics there are significant differences as shown in figs.~\ref{fig_06} and \ref{fig_07}.\smallskip\\
4) In conclusion, it can be added that the temperatures in collisions are estimated as $\omega$ (200 GeV) $>$ $\omega$ (62.4 GeV): $T_0 = 0.7$ GeV (200 GeV) and $T_0  = 0.4$ GeV (62.4 GeV), respectively, provided that the explicit transition (hadrons $\Leftrightarrow$ QCD) (See \cite{Mohanty:2003va}) was not taken into account. In a future we are going to investigate the role of the phase transition between hadrons and QCD in this scheme~\cite{Mohanty:2003va} and problem of the fluctuation~\cite{Biyajima:2004ub}.


\section*{Acknowledgements}
The authors express their gratitude to N.~Suzuki for his collaboration at early stage and useful suggestions for some mathematical problems, and to S.~Muroya for his available comments.


\end{document}